# A SHORT CIRCUIT IN THE ELECTRICAL CABLES WITH POLYMER INSULATION: A NEW LOOK AT THE CAUSE OF ITS OCCURRENCE AND NON-TRADITIONAL WAYS OF SOLVING THE PROBLEM

**Kryshtob VI, Vlasov DV, Mironov VF, Apresyan LA, Vlasova TV, Rasmagin S.I, Kuratashvili ZA, Solovskiy AA**

## Annotation

It is known that most of the electrical cables use as insulation compositions based on polyvinyl chloride. Like most polymers, the latter is quite sensitive to thermal aging, which is not without reason, to be one of the main causes of the various types of faults in the polymeric insulation, leading eventually to a short circuit and fire. On the example of the most common polymer insulator-PVC subjected to preliminary partial thermolysis, simulating the process of accelerated aging, we for the first time show that in this case as a result of the aging process, the electrical conductivity of PVC can acquire abnormal (not obeying Ohm's law) character. In this case, transitions from a state with normal (low) conductivity of PVC into the state with an abnormally high conductivity was clearly observed, being spontaneous uncontrollable process. Especially the large-scale nature of these changes allowing easily transferring polyvinylchloride from a state of typical dielectric (insulator) in a class of conductors attracts attention.). Thus, another one of the most important features of a short circuit in the PVC insulation is opened. It is noted that the reduction of fire hazards in this case (by the maximum possible elimination of the phenomenon of short circuit due to thermal aging of polymeric insulation) should be preventive in nature, as the most effective, simple and convenient way.

**[Короткое замыкание в электрических кабелях с полимерной изоляцией: новый взгляд на причину его возникновения и пути нетрадиционного решения проблемы.**

## Аннотация

Известно, что для большей части электрических кабелей в качестве полимерной изоляции используют композиции на основе поливинилхлорида. Как и большинство полимеров, последний достаточно чувствителен к тепловому старению, которое не без оснований считают одной из главных причин возникновения различного рода неисправностей в полимерной изоляции, приводящих со временем к короткому замыканию и пожару. На примере наиболее распространенного полимерного изолятора- поливинилхлорида, подвергнутого предварительно операции частичного термолиза, имитирующего процесс его ускоренного старения, нами впервые показано, что в этом случае в результате процесса старении значения электропроводности поливинилхлорида могут приобретать аномальный (не подчиняющийся закону Ома) характер. При этом отчетливо наблюдаются переходы поливинилхлорида из состояния с обычной (низкой) проводимостью в состояние с аномально высокой проводимостью, носящие спонтанный неуправляемый характер. Особо обращает на себя внимание масштабный характер этих изменений, позволяющий легко переводить поливинилхлорид из состояния типичного диэлектрика (изолятора) в класс проводников. Таким образом, вскрыта еще одна из важнейших особенностей возникновения короткого замыкания в изоляции из поливинилхлорида. Отмечено, что способ устранения пожарной опасности в этом случае (путем максимально возможного устранения явления короткого замыкания по причине теплового старения полимерной изоляции) должен носить превентивный характер, как наиболее эффективный, простой и удобный.

**Ключевые слова**: полимерная изоляция электрических кабелей, тепловое старение, короткое замыкание, аномалии электропроводности.

## Введение

Одной из основных причин возникновения пожарной опасности в электрических кабелях с полимерной изоляцией принято считать ее тепловое старение, возникающее, например, при неполных коротких замыканиях, перегрузках в электросетях, поверхностных токах утечки и т. д. [1].

До сих пор принято было считать, что под воздействием температурного поля в зависимости от его интенсивности и времени экспозиции, например, на полимерную изоляцию из ПВХ, происходит испарение пластификатора, окисление и деструкция смолы, вызывающие ухудшение комплекса основных физико-механических свойств полимерной изоляции, которое выражается в потере высокоэластических свойств, появлении хрупкости, нарушении ее механической целостности и т. д., что, собственно, принято считать сегодня одной из основных причин проявления короткого замыкания, и как следствие ухудшения пожаробезопасных свойств изоляции. Поэтому в кабельные ПВХ-пластикаты, как правило, сегодня вводят различного рода стабилизаторы и добавки, цель которых предотвращать процессы деструкции, возникающие в процессе как переработки пластикатов, так и эксплуатации кабельных изделий из них [2].

Тем не менее, работы последних лет, связанные с использованием в ПВХ новых типов пластификаторов [3-5] заставили по новому взглянуть на возможность безопасного использования ПВХ в качестве полимерной изоляции в электрических кабелях. Дело в том, что при использовании в ПВХ-пластикатах новых типов пластификаторов впервые удалось зафиксировать совершенно неожиданные, носящие аномальный характер электрофизические свойства, связанные, например, с различными механизмами переключения электропроводности. Поэтому понятно, что весьма целесообразно было изучение влияния на изменение электрофизических свойств самого ПВХ в условиях, моделирующих его тепловое старение (но без использования при этом каких-либо пластификаторов, специальных добавок, ионизирующих излучений и т. д.).

## Экспериментальные данные и их обсуждение

Как указывалось выше, в работах [3-5] при использовании новых типов пластификаторов пленок ПВХ и исследовании их электропроводящих свойств авторами данной работы впервые были обнаружены эффекты сильной нелинейности, релаксационные дрейфы, а также спонтанные обратимые переходы между состояниями низкой и высокой проводимостью.

Приведем кратко наиболее важные полученные результаты. На рис. 1 представлена схема измерительной установки, позволившей проводить исследования ПВХ-пленок [3-5].

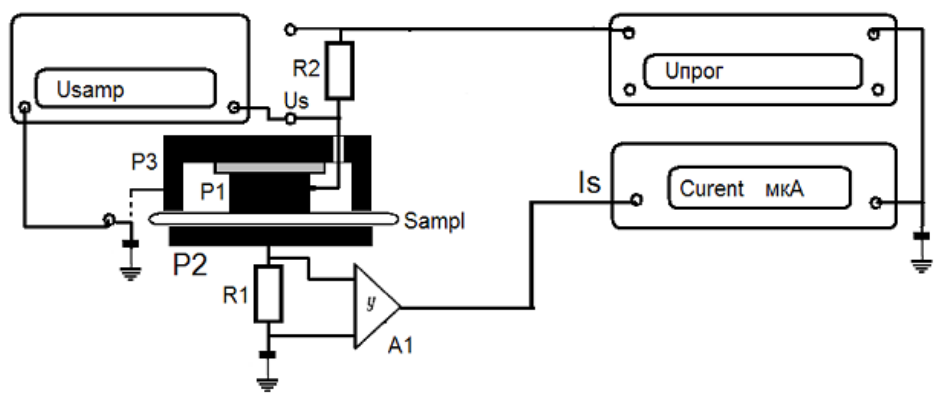

**Рис. 1.** Схема экспериментальной установки для измерений электрофизических параметров полимерных пленок: $U_{samp}$ – напряжения на образце и $I_s$ – тока через образец, в ГОСТированной кольцевой ячейке, состоящей из трех электродов: центральный (P1-центральный электрод диаметром 25 мм, P2 – общий электрод, P3 – электрод для измерения поверхностного сопротивления (P1-P3).

Исследования проводились в области напряжений как выше, так и ниже порога пробоя, не приводящего при этом к разрушению полимерной пленки, поэтому эффекты могли наблюдаться многократно; более подробно особенности измерения электропроводности описаны ранее [5]. На рис. 2 отчетливо видны переходы значений электропроводности

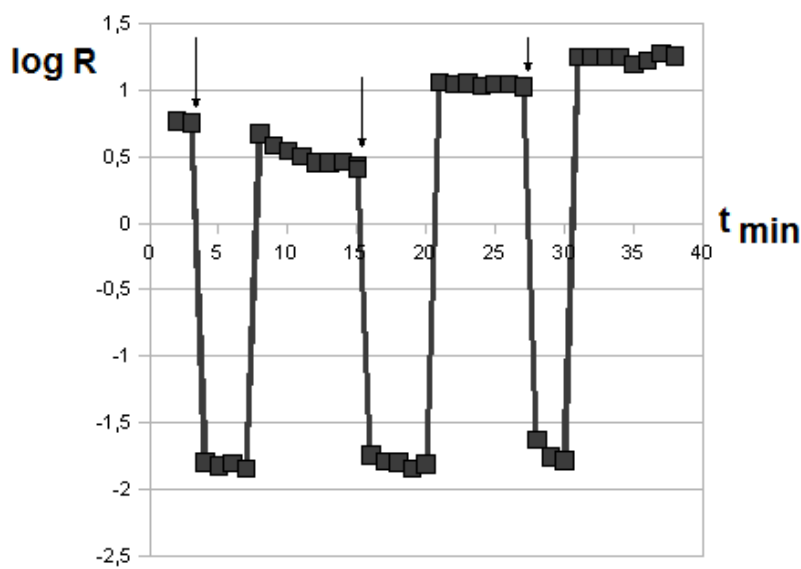

**Рис. 2.** Зависимость сопротивления образца (в логарифмическом масштабе) от времени. Стрелками на графике обозначены точки приложения импульсных перепадов напряжения [5].

ПВХ-пленки (с использованием нового пластификатора) из состояния низкой проводимости (СНП) в состояние с высокой проводимостью (СВП). При этом СВП оказывается квазиустойчивым (время жизни СВП не превышает 10 мин), после чего образец спонтанно переходит в исходное состояние низкой проводимости (СНП).

На рис. 3 приведена зависимость напряжения непосредственно на образце и отношение напряжения к току (в линейной системе эквивалентного сопротивления) от общего

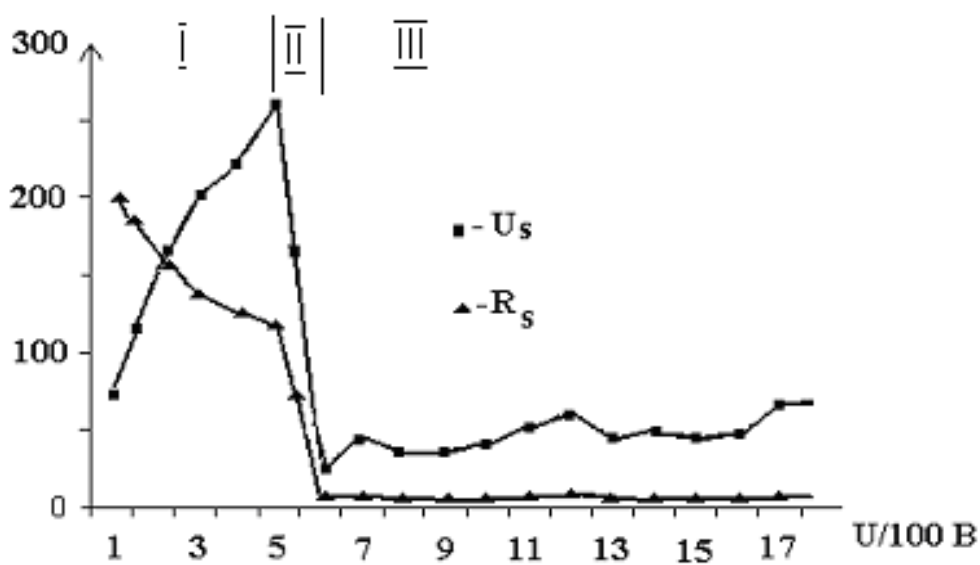

**Рис. 3.** Зависимости отношения напряжения к току образца ($R_S$) в произвольных единицах и напряжения на пленке ПВХ U (В) от общего напряжения на ячейке: ■ – напряжение, ▲ – вычисленное отношение напряжения к току, т. е. наблюдаемое значения сопротивления пленки ПВХ при напряжениях ниже и выше порога перехода в СВП [5].

приложенного к системе напряжения программируемого источника в диапазоне ниже и выше порога «мягкого» пробоя ПВХ-пленки (толщиной 100 мкм). Здесь можно выделить три характерных участка.

I – область плавного уменьшения сопротивления образца, где его эффективное сопротивление нелинейно зависит от приложенного напряжения.

II – область перехода в СВП или область «мягкого» пробоя, в которой сопротивление образца становится существенно меньше балластного сопротивления и напряжение на образце падает, поскольку приложенное к измерительной ячейке напряжение источника практически полностью «садится» на балластном сопротивлении.

III – область СВП, в которой могут наблюдаться сильные флуктуации напряжения на образце.

Таким образом, в работе [5] впервые были обнаружены и исследованы аномалии электропроводности пластифицированных ПВХ-пленок (толщиной до 100 мкм) при напряженности поля как выше, так и значительно ниже порога пробоя. Было также установлено, что аномалии проявляются в виде спонтанных обратимых переходов между двумя состояниями (СНП и СВП), отличающимися на четыре, и более порядков [5].

С другой стороны, хорошо известно, что тепловое старение ПВХ сопровождается явлением дегидрохлорирования (т. е. отщеплением от его макромолекул хлористого водорода), ведущего не только к существенному изменению комплекса его основных физико-механических свойств [6], но предположительно и к изменению его электропроводящих свойств. Ведь хорошо известно, что полиацетилен (ПА), условно представляющий собой ПВХ, подвергнутый операции 100% дегидрохлорирования, обладает значительно меньшим (на 6-8 порядков) показателем удельного объемного сопротивления, чем ПВХ [7].

Учитывая все это, было решено исследовать изменение электропроводящих свойств самого ПВХ, подвергнутого операции частичного дегидрохлорирования, имитирующего процесс его ускоренного старения. С этой целью исследовались модифицированные образцы ПВХ, полученные указанным методом без использования при этом каких-либо пластификаторов, специальных добавок, ионизирующих излучений и т.д.

Данные по электрофизическим и технологическим свойствам полученных в лабораторных условиях образцов исходного ПВХ и его частично дегидрохлорированных образцов представлены в таблице 1 и рис. 4.

**Таблица**

| № образца | Время дегидрохлорирования, мин | Толщина пленок, мкм | Внешний вид и органолептические свойства | Поперечное (объемное) сопротивление, $R_V$, Ом | | Продольное (поверхностное) сопротивление, $R_S$, Ом | |
|---|---|---|---|---|---|---|---|
| | | | | СНП | СВП | СНП | СВП |
| (1) | (2) | (3) | (4) | (5) | (6) | (7) | (8) |
| №1 (исх. ПВХ) | 0 | 14 | Бесцветная, прозрачная, хрупкая, ломающаяся пленка | $>10^{12}$ | Переход в СВП не набл. | $>10^{12}$ | Переход не наблюдается |
| №2 | 20 | 12 | Прозр., менее хрупкая пленка слабо-желтого цвета | $>10^{12}$ | Переход в СВП не набл | $>10^{12}$ | Переход не наблюдается |
| №3 | 240 | 10 | Прозр., желтого цвета, прочная, незалипающая, гнущаяся, легко снимающаяся со стекл.подл. пленка | $8,6 \cdot 10^{11}$ | 1,3 | $>10^{12}$ | Переход не наблюдается 2,2 $10^4$ |
| №4 | 320 | 10 | Прозр., желтого цвета, прочная, незалипающая, гнущаяся, легко снимающаяся со стекл.подл. пленка | $7,1 \cdot 10^{11}$ | 0,6 | $>10^{12}$ | Переход не наблюдается 1,8 $10^{11}$ |
| №5 | 480 | 11 | Прозр., более насыщенного желтого цвета, незалипающая, легко снимающаяся со стекл.подл., гнущаяся пленка | $4,3 \cdot 10^{11}$ | 0,5 | $>10^{12}$ | Переход не наблюдается |

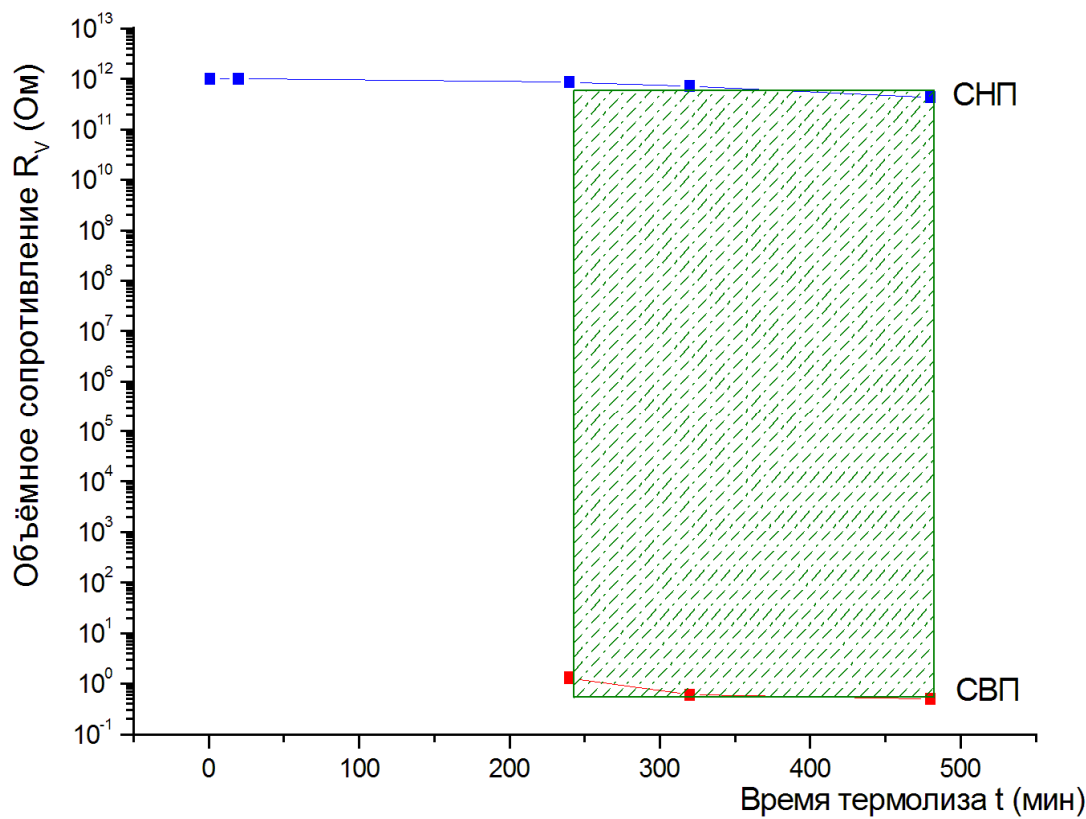

**Рис. 4**. Зависимость объемного сопротивления $R_v$ (Ом) образцов ПВХ от времени термолиза (СНП-состояние низкой проводимости, СВП-состояние высокой проводимости; заштрихована область аномальных спонтанных обратимых переходов).

Наиболее впечатляющим оказалось то обстоятельство, что для модифицированных образцов (обр. № 3-5) переход из СНП в СВП носит столь масштабный характер (более 12 порядков), что образец из состояния типичного диэлектрика легко и непредсказуемо спонтанно может переходить в разряд проводников (со всеми вытекающими отсюда последствиями при использовании в электрических кабелях с полимерной изоляцией).

Обратим внимание на то, что переход в СВП частично дегидрохлорированных образцов ПВХ осуществлялся гораздо легче и быстрее в случае большего времени их термолиза (т. е. в условиях более глубокого процесса его теплового старения).

Необходимо также отметить и то обстоятельство, что близкие к металлическим измеренные значения сопротивления R при использовании стандартной формулы

$$R = \rho L / S \qquad (1)$$

где $\rho$ – удельное объемное сопротивление, L – толщина образца, S – площадь электрода, дают в пересчете на геометрию $R_S$ сопротивления порядка кОм. Однако в эксперименте это сопротивление превышает $10^{12}$ Ом. Эта величина однозначно свидетельствует о том, что расчет удельных сопротивлений и применение соответствующей формулы пересчета (1) для ПВХ-образцов не имеет физического смысла даже для оценок порядка величины.

Поэтому, с учетом всего вышесказанного, основной задачей при использовании электрических кабелей с полимерной изоляцией из ПВХ следует считать исключение ко-
6

роткого замыкания, в первую очередь по причинам ее теплового старения (например, в процессе ее изготовления, перегрузках с сети, превышения установленных норм, времени и условий эксплуатации и т. д.)

Однако не стоит забывать, что устранение КЗ по причинам теплового старения изоляции – это лишь часть «айсберга» решения в целом проблемы обеспечения пожарной безопасности, так как наряду с устранением КЗ по прежнему остается проблема дальнейшего усиления огнезащитных свойств самой полимерной изоляции (горючести, распространения пламени, индекса кислородного горения и т. д.). Кроме того, как никогда остро стоит проблема предельно возможного улучшения одновременно таких показателей, как «Токсичность» и «Дымы».

Достижение всех этих целей одновременно может лежать лишь в рамках использования превентивного, а потому доступного, удобного, высокоэффективного, носящего всеобъемлющий характер, способа обеспечения пожарной, экологической и санитарно-химической безопасности полимерных материалов на всем пути их движения: от производства, эксплуатации включая пожары) до утилизации (посредством обычного сжигания). Такой способ решения высказанных проблем впервые был разработан в России [7] и успешно опробован и внедрен в промышленное производство [8].

**Заключение**

В итоге можно сделать следующие основные выводы.

1. Впервые обнаружены аномальные проявления электропроводности ПВХ-пленок, подвергнутых ускоренному тепловому старению, которые сопровождаются переходами из СНП в СВП, носящими спонтанный обратимый характер.

2. Масштаб изменений электропроводности при переходе из СНП в СВП находится в столь гигантском диапазоне, что позволяет образцам легко переходить из состояния диэлектрика в состояние проводника, с чем нельзя не считаться при использовании электрических кабелей с полимерной изоляцией.

3. Описанные выше аномалии проводимости образцов ПВХ-пленок, не подчиняющейся закону Ома, с проявлениями сильной нелинейности в зависимости от толщины образца, приложенного напряжения, возможности спонтанных переходов в СВП, приводят к потере смысла в расчетах удельной проводимости, измеряемой по ГОСТ. Это требует обязательного учета эффективной проводимости (сопротивления) изоляции из полимеров при проведении измерений, хотя бы с целью существенного повышения точности самого измерения.

4. Для устранения явления короткого замыкания в электрических кабелях с полимерной изоляцией по причине ее теплового старения предлагается разработка нового нетрадиционного способа (носящего исключительно превентивный характер) и являющегося при этом составной частью ранее разработанного всеобъемлющего способа по обеспечению пожарной и экологической безопасности полимерных материалов на всем пути их движения от производства и эксплуатации (включая пожары), до безопасной утилизации (методами обычного сжигания).

**Список литературы**.

1. Г.И.Смелков. Пожарная безопасность электропроводок. М.: ООО «Кабель», 2009.

2. К.С.Минскер, Г.Т.Федосеева. Деструкция и стабилизация поливинилхлорида. М.: Химия, 1972.

3. V.I.Kryshtob, L.A.Apresyan, D.V.Vlasov, T.D.Vlasova.
Anomalous effects in conductivity of thin film samples obtained by thermolysis of PVC in solution ArXiv 1211.3593.

4. Д.В.Власов, Л.А.Апресян, Т.В.Власова, В.И.Крыштоб.
Нелинейный отклик и два устойчивых состояния электропроводности в пластифицированных прозрачных поливинилхлоридных пленках
ПЖТФ, 2010, том 36, выпуск 19, с. 100-107.

5 Д.В.Власов, Л.А.Апресян, Т.В.Власова, В.И.Крыштоб.
Аномалии и пределы точности измерений электропроводности в пластифицированных Прозрачных поливинилхлоридных пленках
Высокомолекулярные соединения, Серия А, 2011, том 53, № 5, с. 739.

6. К.С.Минскер Структурно-физическая стабилизация поливинилхлорида в растворе.
Химия и компьютерное моделирование. Бутлеровские сообщения. 2001,
№ 4, приложение номера.

7. В.И.Крыштоб, Л.А. Апресян, В.Ф.Миронов, Т.В.Власова.
Полимерные материалы: экологическая, санитарно-химическая и пожарная безопасность в условиях их производства, эксплуатации и утилизации. XI международная научно-практическая конференция «Фундаментальные и прикладные исследования, разработка и применение высоких технологий в промышленности», 27-29 апреля 2011 Санкт-Петербурге, Россия,

8. www.specstrol.ru